# Suppression of the Spectral Weight of Topological Surface States on the Nanoscale via Local Symmetry Breaking

*Omur E. Dagdeviren,*[1,2,†] *Subhasish Mandal,*[1,3,†] *Ke Zou,*[1,3,†] *Chao Zhou,*[1,2] *Georg H. Simon,*[1,2] *Frederick J. Walker,*[1,3] *Charles H. Ahn,*[1-4] *Udo D. Schwarz,*[1,2,5] *Sohrab Ismail-Beigi,*[1,4] *and Eric I. Altman*[1,5*]

[1]Center for Research on Interface Structures and Phenomena (CRISP), Yale University, New Haven, Connecticut 06520, USA

[2]Department of Mechanical Engineering and Materials Science, Yale University, New Haven, Connecticut 06520, USA

[3]Department of Applied Physics, Yale University, New Haven, Connecticut 06520, USA

[4]Department of Physics, Yale University, New Haven, Connecticut 06520, USA

[5]Department of Chemical and Environmental Engineering, Yale University, New Haven, Connecticut 06520, USA

# Abstract

In topological crystalline insulators the topological conducting surface states are protected by crystal symmetry, in principle making it possible to pattern nanoscale insulating and conductive motifs solely by breaking local symmetries on an otherwise homogenous, single-phase material. We show using scanning tunneling microscopy/spectroscopy that defects that break local symmetry of SnTe suppress electron tunneling over an energy range as large as the bulk band gap, an order of magnitude larger than that produced globally via magnetic fields or uniform structural perturbations. Complementary *ab initio* calculations show how local symmetry breaking obstructs topological surface states as shown by a threefold reduction of the spectral weight of the topological surface states. The finding highlights the potential benefits of manipulating the surface morphology to create devices that take advantage of the unique properties of topological surface states and can operate at practical temperatures.

**I. Introduction**

Topological insulators (TIs) have gained much attention due to the unusual nature of their electronic surface states (SS), which are protected against standard electronic perturbations largely by time-reversal symmetry [1-4]. Similarly, topological crystalline insulators (TCIs) have SS that are protected by crystal symmetry along particular directions: SS exist in the form of massless Dirac fermions while the bulk is semiconducting [5-8]. Tin telluride, a heavily studied material due to its electrical, optical, and thermal properties, is a TCI up to at least room temperature [8-11] (*see* Ref [12] *for details*). Gapless SS have been observed on (001) and (111) SnTe facets, which have rotational symmetry [6, 8-10, 13]. Global perturbations of the SnTe structure that maintain the crystal symmetry shift the Dirac point in momentum space but keep the surface gapless, [10, 14] while asymmetrically straining the crystal perturbs the SS and opens a global gap [15-17].

We use scanning tunneling microscopy/spectroscopy (STM/S) and density functional theory (DFT) to investigate how local symmetry breaking can be used to manipulate the spectral weight of the topological states of a TCI. Our experimental results show that symmetry-breaking defects suppress electron tunneling, spatially spanning nanometer length scales and energetically ranging over the width of the bulk bandgap at room temperature with no external fields. Complementary *ab initio* calculations confirm the suppression of the spectral weight of topological states and imply that local symmetry breaking may lead to coexisting massive Dirac fermions together with massless Dirac fermions. Decreasing the densities of states by locally perturbing the surface structure is alluring as it permits the patterning of insulating and conducting regions in the same material, e.g., via a scanning probe tip or annealing in vacuum,[18] essential to forming devices. Moreover, the restricted scattering and spin-locked nature of the topological SS [9] opens horizons for new types of transistors, quantum-dots, microwave circuits, and spin-orbit based devices [19-21].

## II. Methods

### A. Experimental

Tin telluride films were grown on SrTiO$_3$ (001) since prior work showed that growth on SrTiO$_3$ yielded SnTe surfaces with regular defects that could be exploited to investigate the effect of symmetry breaking defects on the surface states of TCIs [22]. The substrates were undoped, single-crystalline SrTiO$_3$ (001) samples supplied by CrysTec GmbH, Germany that had been etched at the company with buffered HF solution to yield single-terminated surfaces [22-24]. Once in our laboratory, the SrTiO$_3$ substrates were annealed in ultra-high vacuum for 30 minutes at 1000 K to remove carbon contamination induced by sample transfer through air [24, 25] while remaining non-conducting for any practical scanning tunneling microscopy experiment [24]. The subsequent SnTe film growth was carried out in background pressures below $1\times10^{-9}$ mbar using a molecular beam epitaxy system detailed elsewhere [22]. Single-crystalline SnTe (99.995%) supplied by Alfa Aesar was used as source material for epitaxial growth and was thermally sublimed for 30 minutes from an effusion cell at 820 K. The growth rate was monitored by a quartz microbalance and was held at 2-unit cells (1.26 nm) per minute. The SrTiO$_3$ substrate was maintained at 670 K during SnTe deposition. Following growth, the film was capped with 20 nm of amorphous Se to protect the sample during sample transfer through ambient conditions to the vacuum systems in which the surface analysis was carried out.

After the sample had been re-introduced into vacuum, the Se-capped SnTe film was annealed at 525 K with a background pressure of ~1×10$^{-11}$ mbar for 10 - 30 min to decap the Se layer. The temperature was measured with a thermocouple attached next to the sample holder. Following this process, LEED patterns such as the one in Figure S.I.1 revealed that the film exposed SnTe (001) surfaces while Auger electron spectra (cf. Figure S.I.2 for an example) confirmed that the Se was entirely removed.

With clean samples at hand, STM experiments were carried out in UHV and at room temperature using two distinct systems, details of which have been given elsewhere [26-28]. One system was used to collect wider range images with electrochemically etched W tips at tunneling currents up to 0.5 nA while the other was used for atomic resolution imaging and tunneling spectroscopy with Pt-Ir tips and tunneling currents between 6 - 50 pA. The tunneling spectra were collected by recording current and finding the numerical derivative of tunneling current. The set point prior to all tunneling spectroscopy experiments was 30 pA with a -200 mV bias voltage applied to the sample with respect to the tip (hence, changes in the tunneling set point cannot account for the observed differences in the tunneling spectra). Throughout the paper, we use the convention that negative sample bias voltages reflect electron tunneling from the sample to the tip.

**B. Computational**

Density functional theory (DFT) calculations for SnTe were performed within the generalized gradient approximation of Perdew-Burke-Ernzerhof for exchange and correlation [29] using the projected augmented wave (PAW) approach as implemented in the QUANTUM ESPRESSO software [30]. Fully relativistic pseudopotentials were employed for both Sn and Te atoms. The plane wave kinetic energy cutoff was set to 30

Ry with a corresponding charge density cutoff of 300 Ry. The theoretically optimized bulk SnTe lattice constant was 6.375 Å, which matches very well the experimental value of 6.35 Å. The bulk band gap with spin-orbit interactions included was comparable to the literature value of 0.19 eV and the gap was found to change as a function of lattice constant similar to literature [8]. A supercell slab method was applied to calculate the band structures and density of states of the surface states. Atoms in the slabs were fixed to their bulk positions.

For all the slab calculations in this work, a minimum of 25 Å of vacuum was included in the simulation cell along (001) to isolate periodic copies. For the stepped structures, the Brillouin zone was sampled by a uniform 4×4×2 mesh for self-consistent calculations and a 10×7×2 mesh for computing the density of states (DOS). For defect-free surfaces, 6×6×1 and 20×20×1 meshes of $k$ points (for a 1×1 surface unit cell) were used for self-consistent and DOS calculations, respectively. The Brillouin zone integration was smoothed via the Gaussian smearing technique with a smearing width of 5 mRy (0.068 eV); the same value was used when computing local or projected densities of states.

**III. Results**

This work focuses on thick tin telluride films (400 unit-cells or ≈ 253 nm) exposing (001) surfaces; at this thickness, the film surfaces exhibit the same structure as the (001) surface of bulk SnTe with X-ray diffraction showing no evidence of strain [18]. Figure 1 (and Figures S.I. 1 and 4) show STM images of (001) flat terraces as large as 10000 nm$^2$ separated by half-unit-cell high steps and free of defects other than Sn vacancies common on this material that shift the position the position of the Fermi level but do not impact TCI properties [6]. Complementary Auger electron spectroscopy experiments (Ref. [18] and Figure S.I 2), together with our STM images, reveal no evidence of surface contamination.

Thus, the experiments focus on the clean (001) surface of bulk SnTe known to exhibit TCI properties [8-10].

Although the wide terraces are atomically flat with large apparently perfect regions, defects are evident over greater length scales: e.g., region I in Figure 1a shows a screw dislocation, and region II threading-type screw dislocations that initiate and terminate on the surface. While screw dislocations in TIs do not disrupt topological SS and can even promote the formation of 1D protected states in the bulk [31], they break the $C_4$ rotational symmetry required for gapless TCI SS and thus may be expected suppress the topological SS on SnTe. In addition, ultra-high vacuum annealing which induced slow, partial sublimation of the film created periodic arrays of pits nucleated at dislocations introduced by small angle boundaries in the epitaxial films (See S.I for more information about the pits) [18]. While defects are usually avoided when studying intrinsic materials properties, here we use the surface defects as a platform to reveal their effect on local topological surface properties.

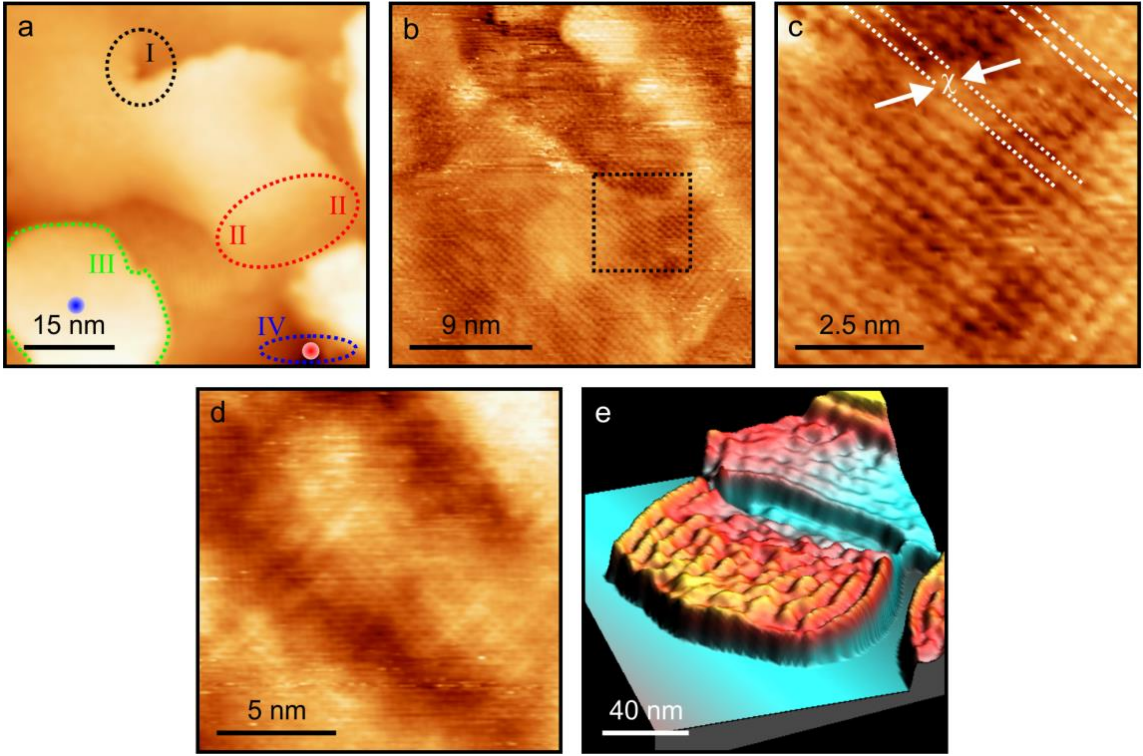

**Figure 1:** Scanning tunneling microscopy images of a SnTe film grown on SrTiO$_3$. (**a**) Overview image ($U_{bias}$ = 2.2 V) demonstrates atomically flat terraces with half-unit cell step heights (3.2 Å). The ellipses highlight: (I) a screw dislocation; (II) paired threading screw dislocations; (III) an island; and (IV) a pit. (**b**) A screw dislocation at atomic resolution (200 mV) showing broken symmetry. (**c**) Zoomed in view of the screw dislocation (200 mV) with the dashed lines and chi showing the half unit cell shift of the atomic rows across the dislocation core, while dashed lines at the corner of the image shows unperturbed lines of atoms (see Figure S.I. 6 for further analysis). (**d**) Atomic resolution image (-100 mV) recorded close to a step edge showing superimposed standing waves that have been reported on topological insulator surfaces [32]. (**e**) Three-dimensionally rendered STM topography showing Standing waves induced by step edges (-750 mV). Valley (black)-to-peak (white) corrugations are 14.81 Å (a), 3.67 Å (b), 2.05 Å (c), 4.66 Å (d), and the base to topmost height is 18.54 Å for (e).

The images in Fig. 1b – d show several of the characteristic defects at higher resolution. In particular, screw dislocations were imaged with atomic resolution (Figs. 1b & 1c), which reveals that the broken local symmetry near the dislocation core laterally shifts the atomic rows by half a surface unit cell (arrows and dotted lines in Fig. 1c). In addition, Fig. 1b, c show obvious contrast differences between even nominally identical surface sites. This background contrast can be ascribed to electron scattering and modification of the local density of states (LDOS) by surface imperfections: Figure 1d displays a periodic perturbation of the heights of the atoms while the longer-ranged image in Fig. 1e shows standing wave patterns emanating from step edges and with a 5.5 nm wavelength and a wave vector perpendicular to the step edges (further details on the standing waves can be found in Ref [12]). Quasiparticle scattering of topological SnTe states in this energy range has been previously reported [33].

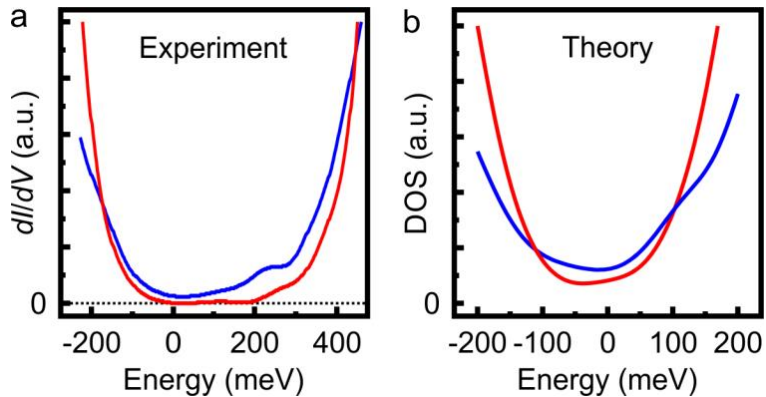

**Figure 2:** (**a**) Tunneling spectra recorded in the middle of a terrace (blue curve recorded within the blue circle in region III of Figure 1a) showing densities of states for a defect-free region, and at an area with high curvature (red curve recorded within red circle in region IV of Figure 1a) revealing suppression of densities of states. (**b**) DFT calculated atom-projected DOS averaged over all surface atoms on the surface layer for infinite terraces (blue) and a stepped vicinal (103) surface

(red) showing suppressed DOS within 50 meV of the Fermi level. These results are for 19-layer thick slabs.

The results in Fig. 1 reveal that defects create electronic contrast in STM images on the 1 – 10 nm length scale. To distinguish the impact of defects on the topological SS, scanning tunneling spectroscopy (STS) experiments were carried out. Conductance (dI/dV) maps recorded as a function of bias voltage at low temperatures are generally used to reveal the local electronic properties [10, 15, 16]. We used a point spectroscopy approach for quantitative analysis of local electronic properties because prior combined STM/NC-AFM measurements on SnTe showed that the tunneling current only becomes measurable when the tip is so close to the surface that the tip–surface interaction is repulsive, resulting in frequent tip changes in the room temperature experiments [18] that impede comparing dI/dV maps collected at many biases. In the middle of a terrace (blue circle in Figure 1a), STS (Figure 2a) indicates no band gap, as expected. Meanwhile, spectra recorded over the steeply sloped area labeled as region IV in Figure 1a (referred to as a "pit" that we typically find surrounds a dislocation created by a small angle boundary as described above [18] and in Ref [12]) reveal a factor of 10 reduction in tunneling conductance implying suppression of the density of states from the Fermi level up to 200 meV [8]. The valence band edge is pinned near the Fermi level due to Sn vacancies that *p*-dope the material [6, 34]. The 200 meV span of the effect is much larger than previously reported intensity shifts due to global symmetry breaking via strain [10, 16]. Figure 3 shows the tunneling spectrum recorded around a different pit, which in this case is part of a periodic array (Figure S.I. 4 a-c) [18]. Figure 3 illustrates how the tunneling conductance (proportional to the LDOS at low bias [35]) increases gradually (spectra at II & III) as the tip moves away from the pit (location

I). The small but finite tunneling conductance around the Fermi level can be associated with thermal broadening in the room temperature measurements as well as the radius of curvature of the tip leading to a contribution to the tunneling current from adjacent conductive areas not directly below the tip, particularly when the tip is positioned within a pit [18]. The edges of the pits are comprised of closely spaced atomic steps and inclined facets that form when the steps merge (see Fig. S.I. 1). Conductance maps recorded with the tunneling bias within the gap region (see Fig. S.I. 1g, h) show limited conductance adjacent to an isolated half-unit cell high step, implying that single atomic steps can also impede the topological surface state.

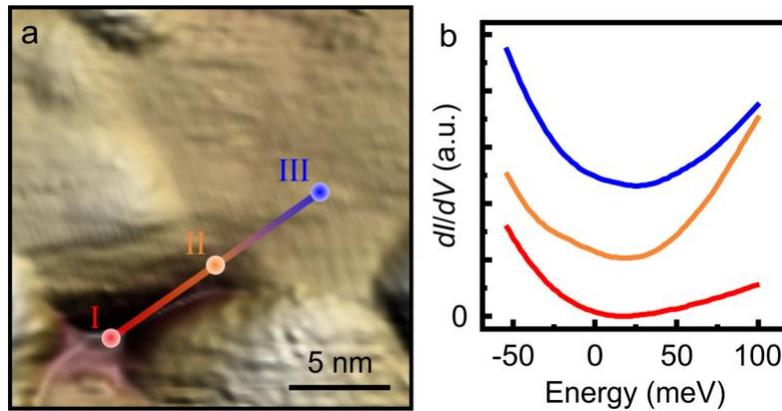

**Figure 3: (a)** Scanning tunneling microscopy topography image of a pit. **(b)** Scanning tunneling spectroscopy experiments conducted along a line. Numerical derivative of tunneling spectrum shows that the tunneling conductance is suppressed within the pit (I), and gradual increases in *dI/dV* (II) are observed as the tip moves to smoother regions (III) where the symmetry of crystal is protected. Note that the vertical scale of the tunneling spectra is amplified 25 times compared to Fig. 2 (See Fig. S.I. 5 for comparison) to show the variation around the Fermi level more clearly. The valley (black)-to-peak (white) corrugation is 3.41 nm for (a).

We consider and rule out several alternative explanations for the reduced local densities of states. The observed gradual increase in the tunneling conductance around the Fermi level (up to ≈50 meV), which implies an increase in the local density of states with distance from the symmetry breaking zone, could possibly be explained by variations in local chemical composition or undetected tip changes. We did not observe clustering of Sn vacancies on the surface that could create tin telluride phases with a Sn:Te ratio less than one or evidence of impurities in Auger electron spectra [18]. Regarding tip changes, we note that the general features of the spectra are reproducible, i.e. the tip can be moved back and forth between the marked locations multiple times without changes.

The fundamental question we wish to address is if the suppression of the conducting surface states can be attributed to defects disrupting the rotational symmetry: to this end we conducted DFT calculations on model systems. The DFT calculations were done with two prototypical surfaces (detail in SI): one is the ideal flat (001) surface preserving symmetry and the other is the stepped (103) surface, which has periodic surface steps along [100] that break the symmetry between the [100] and [001] directions. The vicinal surface used in the computation contains step edges to replicate the experimental situation, and at the same time its periodic nature makes DFT calculations tractable. Figure 2b highlights a qualitative analysis based on the atomic-orbital projected density of states (PDOS) on the surface atoms: the stepped surface has a noticeably reduced density of states near the Fermi level. Other qualitative spectral features reproduced by theory include a hump near 100 meV for the defect-free surface, which corresponds to Sn $p$ dominated states, seen at 200 meV in the experimental spectrum of Figure 2a (the difference in energy is due to DFT underestimating band gaps and the p-doping of the experimental sample) [36].

The agreement between DFT and experiment in Figure 2 is qualitative with a weaker reduction in the PDOS due to symmetry breaking defects; we discuss three reasons for this discrepancy. First, computing the PDOS requires broadening (smearing) of the discrete set of computed energy levels from the finite simulation size, and our results are based on a Gaussian broadening of 70 meV. Second, a PDOS should not be compared directly and quantitatively to the experimental dI/dV: since tunneling happens in real-space, the real-space local density of states (LDOS) is a better quantitative surrogate for dI/dV and shows a much larger reduction near the Fermi level as illustrated in Figure 4. Third, our simulation cells are too small to distinguish differences of electronic structure between step edges and flat terraces in the same calculation: we have simulated 3-, 5-, and 7-atom wide terraces (~1-2 nm wide) and all show reduced surface DOS near the Fermi level (see SI). Given the experimental observation that one must move ~10-15 nm away from a step edge to recover the clean surface DOS (Figure 3 and SI), the required simulation cells to directly reproduce this effect and isolate the effect of a single step edge are too large for present day DFT calculations.

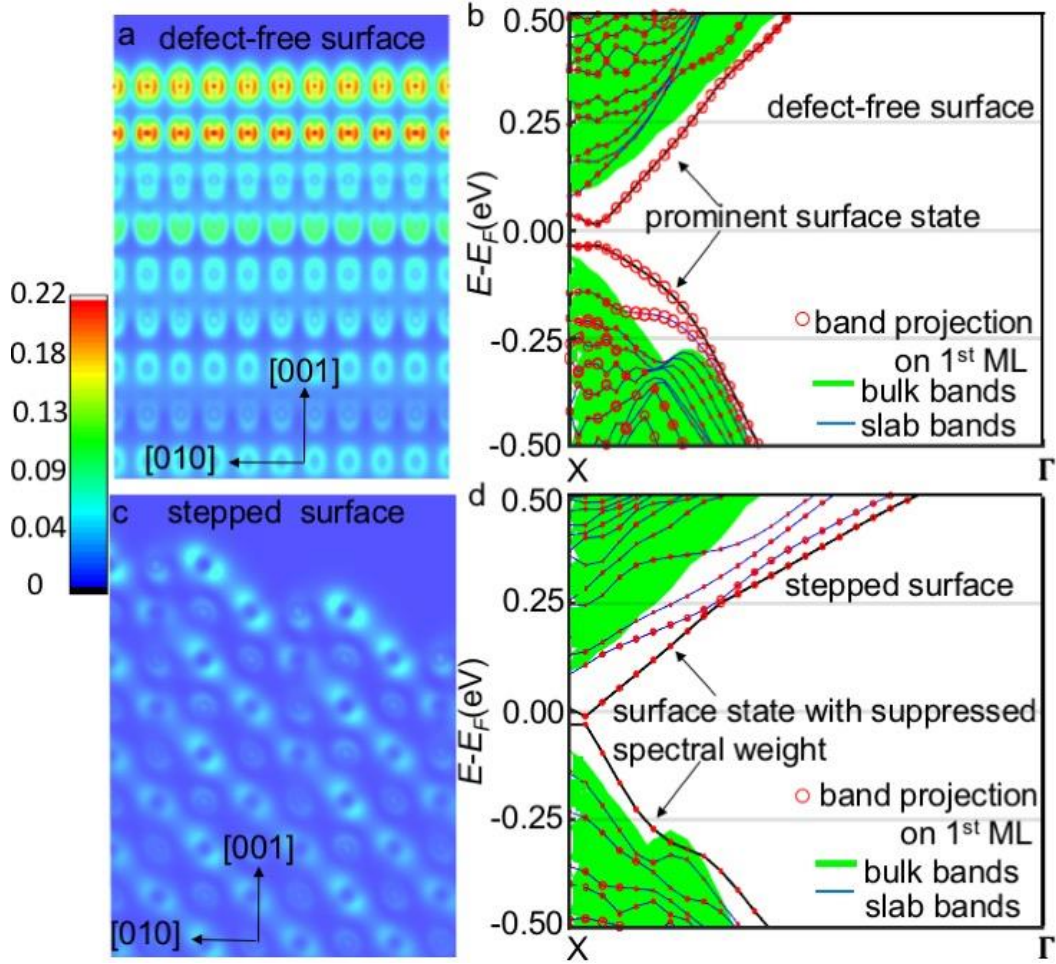

**Figure 4:** DFT calculations for 19 atomic layers thick (001) SnTe slabs. (**a**) & (**c**) Integrated from $E_F$ - 0.1 eV to $E_F$ + 0.1 eV, capturing predominantly the surface states, and averaged (along [100]) LDOS in real-space for the (a) defect-free and (c) the stepped (001) SnTe surfaces. Higher (lower) LDOS is shown by red (blue) colors on a same color scale plot. (**b**) & (**d**) Theoretical band structure for defect-free (c) and stepped surfaces (d); the green-shaded area refers to the bulk bands while blue lines are computed for finite slabs. The SS forming the Dirac point are shown in thick black color with their weight indicated by the size of the red circles (band projection). The SS exists in both cases but the spectral weight of the topological surface states is greatly reduced for the stepped surface. The small gap for the SS in (b) is due to the finite slab thickness [37].

For a better comparison with STM data, Figure 4 shows a comparison of the real-space LDOS for the two model systems on the same scale. The contrast between defect-free and defective surfaces is striking: the LDOS reduction is quite dramatic and is numerically a factor of ~3, which means that the spectral weight of metallic SS have been nearly eliminated by the symmetry-breaking defect. We also note that the atoms protruding at the edges have a much higher LDOS than those in the flat terrace regions of the surface; this chemical effect, due to dangling bonds, is in agreement with how atoms near vacancies appear brighter in the STM images (Fig. S.I.3). This level of detailed agreement gives further confidence regarding the relevance of the modeling.

The DFT band structure of the thick defect-free surface shows the expected Dirac-point near the X-point (Figure 4b and Figure S.I. 7). A very small hybridization gap (< 0.04 eV) appears due to the slab's finite thickness: when the slab thickness is increased from 4 ML to 19 ML, this artificial gap decreases [37] (see Figure S.I.9). The projection of each band on the surface atomic orbitals shows that, as expected, the states near the Dirac point have strong projections on the surface atoms (Figure 4b and Figure S.I. 8). The band structure plot for the symmetry broken stepped surface is provided in Figure 4d. A Dirac point is observable near the Fermi energy, which is due to the existence of flat regions in the simulation, but the surface projections are reduced significantly, indicating the suppression of spectral weight of the topological surface states. These results provide a physical picture of what the band structure of a defective surface with larger flat terraces should look like: the flat regions would support massless Dirac fermions while simultaneously regions near the defects would have greatly suppressed weights for the Dirac bands; if the defective region is wide enough, a gapped spectrum with massive fermions would be observed inside it. Dirac mass generation by opening up a gap due to *globally* breaking the required symmetry over an entire crystal was recently demonstrated for $Pb_{1-x}Sn_xTe$ [15]. Here, we argue that Dirac mass generation can potentially be achieved by locally breaking the rotational symmetry on the surface by creating extended defective regions. Hence, the global band structure or the global Dirac points, e.g., as measured by angle-resolved photoemission spectroscopy (ARPES), can remain gapless even though locally gaps can form. From the ARPES viewpoint, the effect of local gap formation is subtle as it simply reduces the spectral weight near the Fermi level compared to the clean surface (see Figure 4). However, local probes can uncover the change of spectral weight directly and dramatically, as described above.

**IV. Summary**

Experiments and *ab initio* calculations revealed that topological SS in TCIs can be inhibited by local symmetry breaking. The impact of such defects has been exposed experimentally via the observation of suppressed conductance in tunneling spectra recorded within defective regions. Our experimental findings are supported and explained by DFT calculations on a prototypical surface that breaks the SnTe crystal symmetry locally and shows spectral weight suppression near the Dirac point. Our results demonstrate that topological SS may be tailored within regions confined to a few nanometers. Such local modification of topological states can be manipulated by *in-situ* heat treatment of epitaxial films [18]. The collapse of gapless SS to a bulk state may enable the fabrication of novel electronic devices operating near room temperature on a single material simply by patterning islands or pits, thereby bypassing complex issues regarding the interactions and compatibilities of different materials across interfaces.

**Corresponding Author**

*Corresponding author: eric.altman@yale.edu

**Acknowledgments**

We thank Profs. Leonid Glazman and Judy Cha for fruitful discussions. The research was funded by the Focus Center on Function Accelerated nanoMaterial Engineering (FAME) and the NSF (Grant No. MRSEC DMR-1119826) through the Yale Materials Research Science and Engineering Center. S.M. was supported by the NSF SI2-SSI program (Grant No. ACI-1339804). Computational facilities are supported by NSF XSEDE resources through Grant No. TG-MCA08X007 and by the facilities and staff of the Yale University Faculty of Arts and Sciences High Performance Computing Center.

†These authors contributed equally to this work: O.E.D., S.M., K.Z.

O.E.D. performed high-resolution STM and STS experiments. S.M. performed the DFT calculations. O.E.D., G.H.S., and C.Z. conducted large-area STM experiments and macroscopic surface characterization. K.Z. developed the growth process. O.E.D. and E.I.A. designed the experiments. O.E.D. and S.M. analyzed the data. O.E.D., S.M., S.I.B., and E.I.A. drafted the manuscript. C.H.A., E.I.A., S.I.B., U.D.S., and F.J.W. conceived the project.

# Supplemental Material for

# Suppression of the Spectral Weight of Topological Surface States on the Nanoscale via Local Symmetry Breaking

*Omur E. Dagdeviren,[1,2,†] Subhasish Mandal,[1,3,†] Ke Zou,[1,3,†] Chao Zhou,[1,2] Georg H. Simon,[1,2] Frederick J. Walker,[1,3] Charles H. Ahn,[1-4] Udo D. Schwarz,[1,2,5] Sohrab Ismail-Beigi,[1,4] and Eric I. Altman [1,5]\**

[1]Center for Research on Interface Structures and Phenomena (CRISP), Yale University, New Haven, Connecticut 06520, USA

[2]Department of Mechanical Engineering and Materials Science, Yale University, New Haven, Connecticut 06520, USA

[3]Department of Applied Physics, Yale University, New Haven, Connecticut 06520, USA

[4]Department of Physics, Yale University, New Haven, Connecticut 06520, USA

[5]Department of Chemical and Environmental Engineering, Yale University, New Haven, Connecticut 06520, USA

[†]These authors contributed equally to this work

# SnTe Topological Properties at Room Temperature

Prior work indicates that SnTe maintains its topological behavior at room temperature. There is a rhombohedral distortion that sets in below 100 K that breaks the symmetry requirements for a TCI, but nonetheless experiments indicate no difference in the topological behavior measured at 30 K and 130 K [1]. Meanwhile, transport measurements show no indication of a phase transition up to room temperature [2] and band inversion is preserved as proposed by former work [3, 4]. The effect of temperature then comes down to thermal expansion. Theory indicates that a lattice expansion of 1.65% would be required to break the band inversion responsible for the topological behavior [5] while thermal expansion from 0 K to 300 K could account for at most only a 0.6% lattice expansion [6]. Room temperature ARPES measurements on In-doped SnTe showing the band dispersion expected for topological surface states support this view [7]. Indium incorporation p dopes the material but does not change the cubic structure and only marginally affects the lattice constant (In's slightly larger size than Sn would push the material towards the trivial insulating state) [2]. Thus, prior works all indicate that SnTe is a TCI at room temperature.

# Experimental Methods

## Low-Energy Electron Diffraction and Analyses of Pit Formations

We performed low-energy electron diffraction (LEED) experiments to characterize the film and its orientation with respect to substrate. As Figure S.I. 1a shows, LEED data of the substrate does not reveal a strong reconstruction, displaying only faint fractional order spots consistent with a (2×2) reconstruction. Figure S.I. 1b is the LEED data for the 250 nm thick SnTe film. The LEED pattern of the film has a mosaic spread of 15°. The mosaic spread observed for the film implies that the substrate has domain-by-domain growth and individual domains are rotated with respect to each other. Periodic dislocations will occur at the small angle boundaries between the domains as illustrated in Fig. S.I. 1g. As discussed previously [8], annealing the SnTe film so that it partially sublimes resulted in periodic pits separated by distances consistent with the expected dislocation spacing, implying that SnTe vacancy pairs produced by sublimation diffuse to the dislocations to form pits. Consistent with this picture, the lateral size and depth of the pits increased with annealing time until they merged into deep grooves [8]. Out-of-plane rotation between neighboring domains can account for the frequent screw dislocations seen in the films [9]. To further elaborate the topographic structure of the pits, we investigated the height profile across the pits. As Figure S.I. 1 c and d reveal that the pit depth is an integer number of half unit cell steps. The individual steps down could be resolved on the right while on the left the steps appear to have merged into a multi-atom high step that can be considered an inclined facet. The topography of the surface area (Figure S.I. e-f) where we conducted scanning tunneling spectroscopy experiments discloses some steps less than half-unit cell high. Sub half-unit cell steps arise at the ramps of screw dislocations. For cohesiveness, we performed conductance mapping of the sample across a half-unit cell step by oscillating the bias voltage and recording an image of the current variation at the oscillation frequency using a lock-in amplifier which provides dI/dV (Figure S.I. h-i). As demonstrated, the conductance (which is related to the local density of states) within the bulk bandgap is suppressed in the vicinity of the half-unit cell step (red and green areas in Figure S.I. i), while with increasing distance from the step, the conductance recovers (represented by blue in Figure S.I. i).

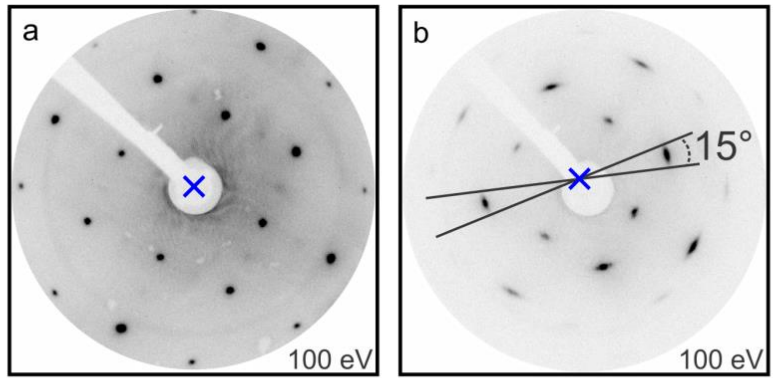
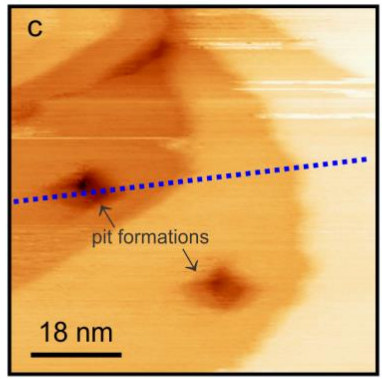
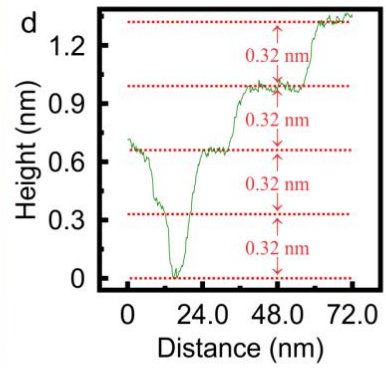
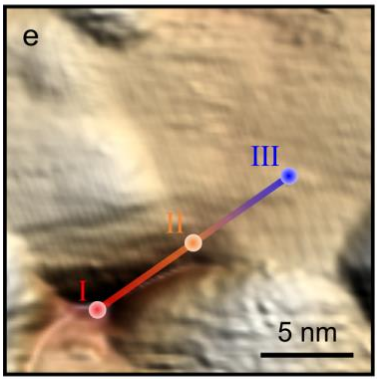
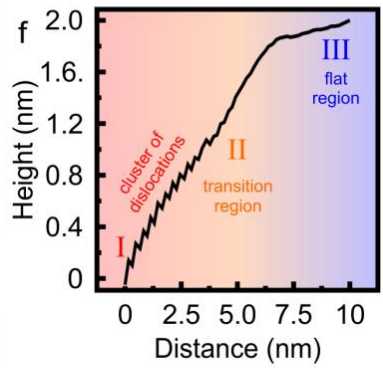
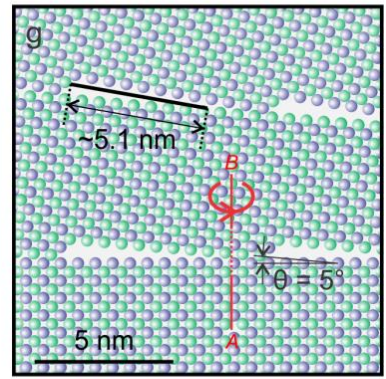
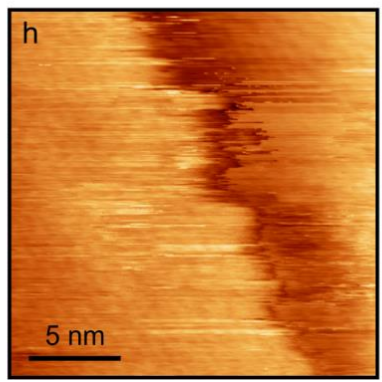
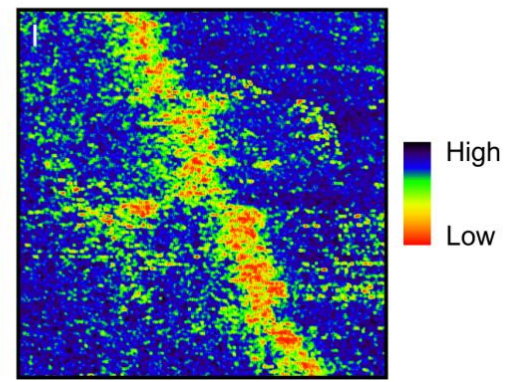

**Figure S.I. 1**: Low energy electron diffraction (LEED) and scanning tunneling microscopy (STM) analyses. (a) LEED of the substrate with a faint (2×2) pattern is shown after the film is sublimed away under UHV conditions. (b) LEED data for a 250 nm thick SnTe film shows that the film has a mosaic spread of growth domains. The arcs indicate a mosaic spread in the film of 15° centered on the substrate [1,1] direction. The 15% compression of the pattern compared to that of the substrate is consistent with an increase in the surface lattice constant towards the 0.447 nm spacing on the SnTe (001) plane. (c) An overview STM image shows the pits that form during the growth (peak to peak corrugation of 1.40 nm). (d) The line profile across the pit shows half-unit cell step heights. The topography and the surface (peak-to-peak corrugation 3.41 nm) where we conducted scanning tunneling spectroscopy experiments (e) and the height profile along the line of spectroscopy experiments (f). Schematic representation of the pits is shown for a small angle domain boundary as adapted from Ref [9] (g) Topography (h) of a half-unit cell step and (h) the corresponding dI/dV map (i) across the step edge ($V_{bias}$=80 mV, $V_{oscillating}$=10 mV at 3 kHz, $I_{set}$=60 pA).

**Auger Electron Spectroscopy**

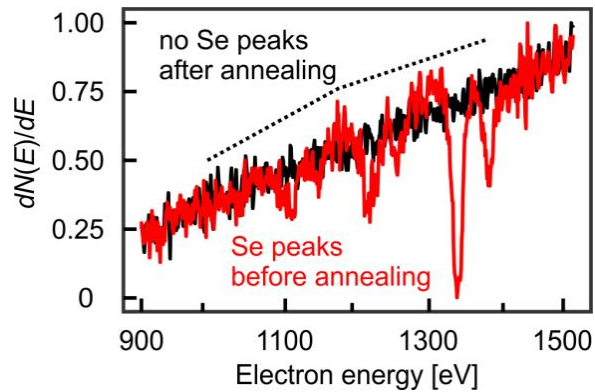

**Figure S.I. 2:** Auger electron spectra of Se-capped SnTe film before (red) and after (black) annealing in ultra-high vacuum at 550±25 K for 10 minutes. Annealing the Se-capped film for 10 minutes at 525 K completely removed the Se. Prior to imaging, the SnTe films were annealed at least 30 minutes in UHV at 525 K to ensure complete Se sublimation. Complete disappearance of Se peaks in the Auger spectrum implies the absence of the Se on the surface. The AES sensitivity factor for these Se peaks is 0.65 [10] .

**Atomic Resolution Characterization of SnTe Film**

An example of an atomic resolution image is provided in Figure S.I.3 which shows atomic defects (indicated by arrows) with a signature consistent with Sn vacancies [11]. At the negative sample bias in this occupied state image the contrast is dominated by Te atoms with Sn vacancies giving rise to apparent distortions in the atomic positions [11].

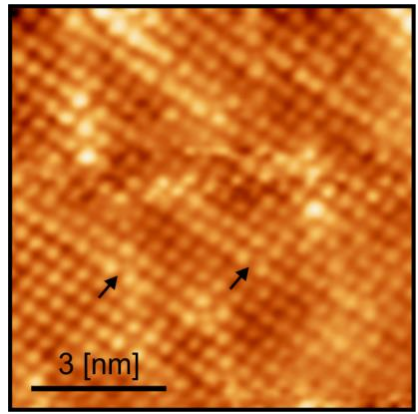

**Figure S.I. 3:** Atomic-resolution scanning tunneling microscopy image of the surface following above-mentioned procedure. Atomic defects in the form of vacancies are indicated by arrows and atoms around the defects appear brighter than the surrounding atoms and displaced outwards [8]. Observed brightness and apparent displacement of atoms are consistent with former experimental observations [11]. Atomic resolution image (-200 mV, 10 pA, 55 pm vertical range).

**Surface Standing Waves**

We observed standing waves on the surface at different sample bias voltages. Figure S.I.4 demonstrates standing waves with different wavelengths, which have been recorded at varying bias voltages. The wavelength of the standing waves attenuated, as the bias voltage was made less negative: from 9 nm at -750 mV (Figure S.I.4) to 4.7 nm at -100 mV (Fig. 1d-e main text). In addition, Variations in the atomic structure of the tip impede quantitative wavelength analysis of the standing waves (Figure 1d-e, Figure S.I. 4); however, the overall trend is still valid.

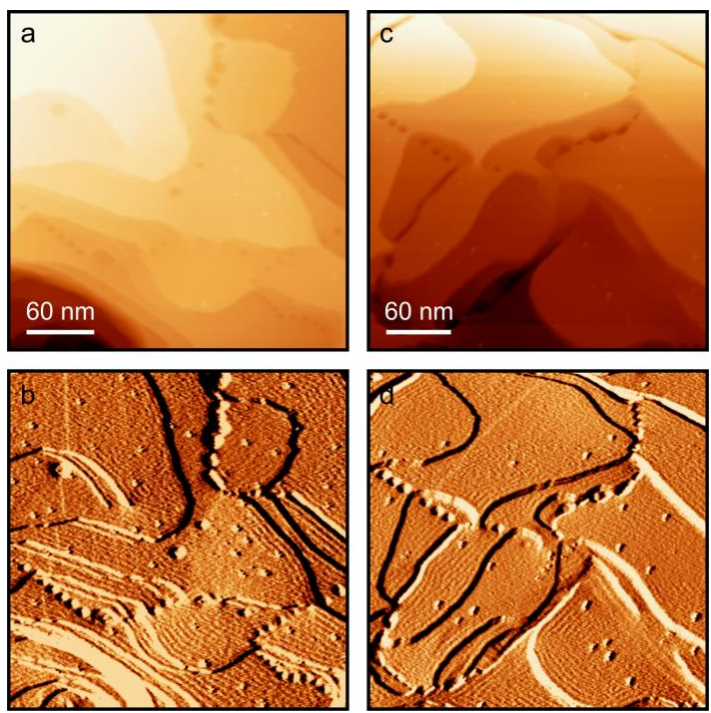

**Figure S.I. 4:** Scanning tunneling microscopy images of standing waves induced by step edges. (a) the surface topography recorded at -750 mV and (b) the derivative of the topography image in (a). The wavelength of the standing waves parallel to the step edges is 5.7 nm. (c) Topography image recorded at -500 mV and (d) the corresponding derivative image. The wavelength of the standing waves is 5.1 nm with a lower height corrugation

than (a), (b). The peak-to-peak corrugation in (a) and (c) are 3.62 and 2.23 nm, respectively.

**Comparison of Scanning Tunneling Spectroscopy Experiments**

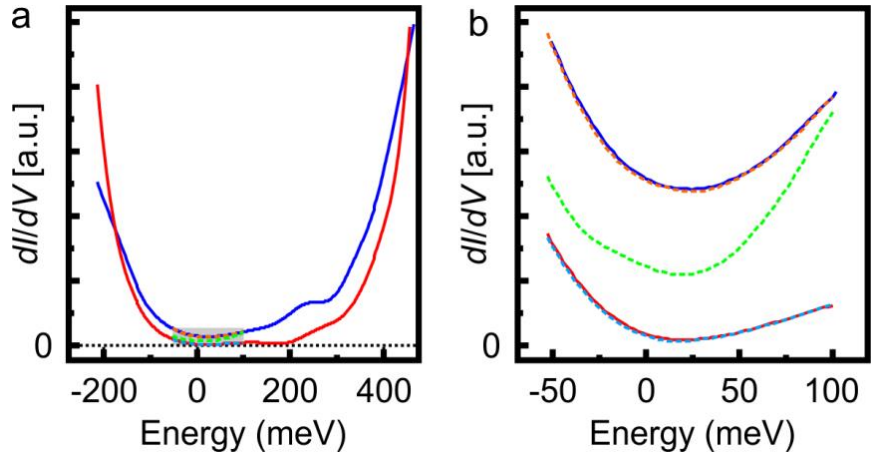

**Figure S.I. 5:** Demonstration of suppression of local densities of states with scanning tunneling spectroscopy (STS) experiments. (a) STS data in Figure 2 (blue and red curves) and Figure 3 (light blue, green and orange dashed curves) in the main text are plotted with the same energy range in Figure 2. (b) The highlighted region in (a) is zoomed in. The agreement in the data demonstrates the reproducibility of STS experiments and reduction of density of states.

**Further Analysis of Screw Dislocation**

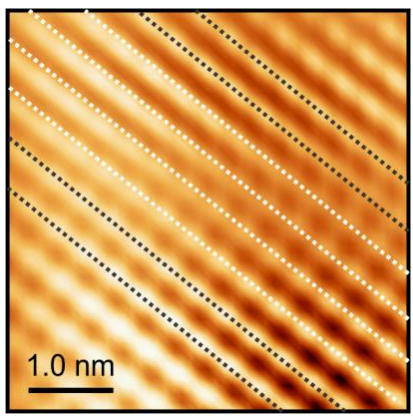

**Figure S.I. 6:** Fourier-filtered image to highlight the screw dislocation in Figure 1c of the main text. Black dashed lines are on top of atoms all the way across the image down and to the left and up and to the right of the screw dislocation. Meanwhile, at the screw dislocation the white dashed lines are directly on top of the atoms on the right of the image but fall between the atoms on the other side of the screw dislocation (up and to the left).

## Computational Methods

Two types of slab structures were considered to investigate gap opening due to crystal symmetry breaking in SnTe: one was a standard flat (001) surface (defect-free) where the crystal symmetry is conserved while the other was a "stepped" structure or (103) vicinal surface where the crystal rotational symmetry was broken locally along the [110] direction while the atomically flat terraces between the steps preserved local crystal symmetry. We constructed regular or defect-free SnTe surfaces along (001) with 8, 18 and 19 monolayers (ML) thick slab. The slabs for the step structures were created with the same thicknesses to distinguish the change in electronic structure only due to the effect of local symmetry breaking. We performed analyses of step structures with 3, 5 and 7 atomic layer wide terraces. The 3 atomic layer wide step structures with 18- or 19-ML thickness slab consist of 104 or 110 atoms in the unit cell, respectively; the 5 atomic layer wide and 8 ML thick SnTe step structure consist of 76 atoms in the unit cell (the 7 ML wide and 8 ML thick unit cell has 108 atoms in the unit cell). Band structures for defect-free SnTe slabs with 18 ML and 19 ML thickness were computed (see Figure S.I. 7). They show the known fact that the Dirac point shifts its location near the X-point slightly depending on whether the SnTe slab contains an odd or even number of layers. The degeneracy at the X point is preserved in the 18 ML slab, protected by two-fold screw rotation symmetry, but is broken in the 19 ML case due to the absence of this symmetry [12]. Band projection for 18 ML thick defect-free slab in a larger window is shown in Figure S.I. 8. The very small gap that appears here as well as in Fig. 4b-d in main text is due to the finite thickness of the slab [13]. To show this hybridization gap as a function of thickness we constructed defect free slabs with 4, 6, 8, 12 ML thickness and plotted the band structures in Fig. S.I. 9 a-d. The gap decreases as the thickness increases as shown in Fig. S.I. 9 e-f. The inverse thickness shows a clear linear behavior with the hybridization gap for these thicknesses.

To demonstrate the robustness of our calculations, we demonstrate the reduction of the surface DOS due to the local breaking of rotational symmetry for stepped vicinal surfaces with 3, 5, and 7 ML wide terraces (the (103), (105) and (107) vicinals) for slabs with thickness of 8ML. In Fig. S.I. 10, we compare the surface layer DOS of these stepped surfaces with that of an infinite terrace (i.e., defect-free surface). We clearly see a strong reduction of the DOS for the stepped surface when compared to a defect-free structure regardless of the terrace width for the sizes under consideration.

Different atoms on a stepped terrace are named edge, edge+1, edge+2: these are the atom on the exposed edge, one atom away going into the terrace, and two atoms away going into the terrace, respectively. Atoms labeled edge-1 are those immediately below the edge atom of the next vicinal step. The projected density of states on different locations on the terrace of a stepped surface with thickness of 8 ML and terrace sizes of 7 ML are show in Fig S.I. 11. We again see the reduction of the DOS for all locations on the terrace when we compared with a defect-free surface. The edge+1 atoms show the strongest reduction. The DOS shown in that figure is averaged over all atoms (Sn and Te) in rows parallel to the step edge. The band structure for this symmetry broken stepped surface is plotted in Figure S.I. 12. A Dirac point is observable near the Fermi energy, which is due to the existence of flat regions in the simulation, but the surface projections are greatly reduced, indicating suppression of the weight of the SS.

In Figure S.I. 13 we show layer by layer weight (probability) of the projection of specific band states as a function of layer number for both the defect-free and stepped surfaces at a k-point near the Dirac point for both valance band maxima (VBM) and conduction band minima (CBM). After fitting these data with a $3^{rd}$ order polynomial, we compute the steepness (f'(x)/f(x) on the first layer where f(x) is the weight on layer x) which is proportional to the decay rate of the surface state going into the interior. The computed steepnesses are similar for both surface structures; -0.39 and -0.37 $ML^{-1}$ for defect-free and stepped surfaces respectively. We note that since each band state is a normalized wave function, the reduced weight of the SS on the surface means that more of its weight must be located in the interior of the slab for the stepped surface.

Figure S.I. 14 and S.I. 15 show top and side views of the unit cell of the calculations reported in the main text. The figures show how the ideal flat (001) surface has the required

two rotational symmetries preserving TCI properties while the stepped simulated structure breaks both rotational symmetries.

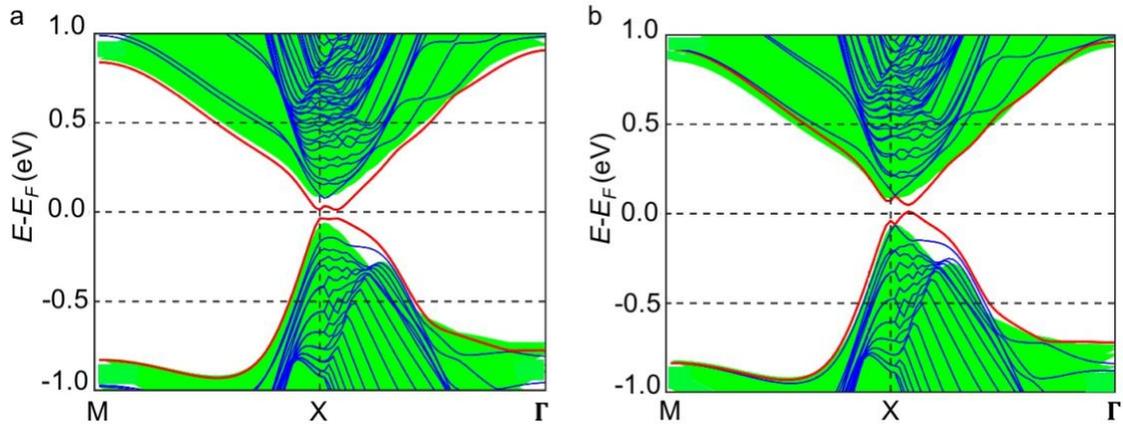

**Figure S.I. 7:** Band structure of defect-free SnTe for (a) 19 ML (odd number) and (b) 18 ML (even number) thick slabs. The green-shaded areas refer to the bulk bands while the blue curves are those computed for the slabs. The SS are highlighted by the red lines that show the Dirac point near the X-point. The Dirac point shifts slightly near the X-point depending on whether slab contains even or odd number of monolayers. The very small gap that appears here, as well as in Fig. 4 in main text, is due to finite thickness of the slab [13].

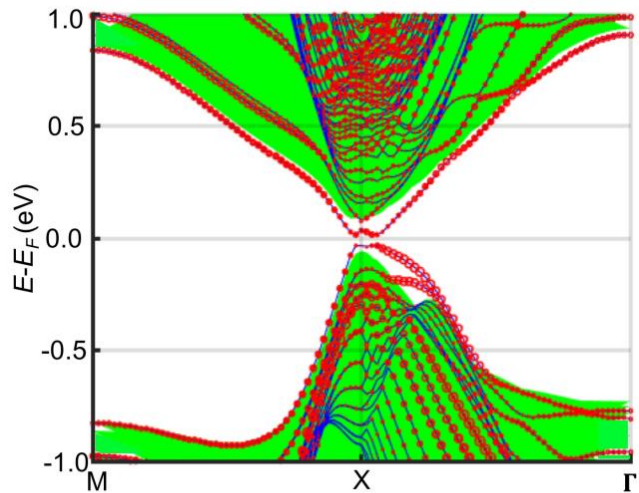

**Figure S.I. 8:** Band structure of defect-free 18 ML thick SnTe slab with projections onto all surface atomic orbitals shown as red circles. The green-shaded areas refer to the bulk

bands while blue curves are those computed for finite slabs. The surface states form the Dirac point at X-point. The weight of the band projection for surface atoms is indicated by the size of the red circles.

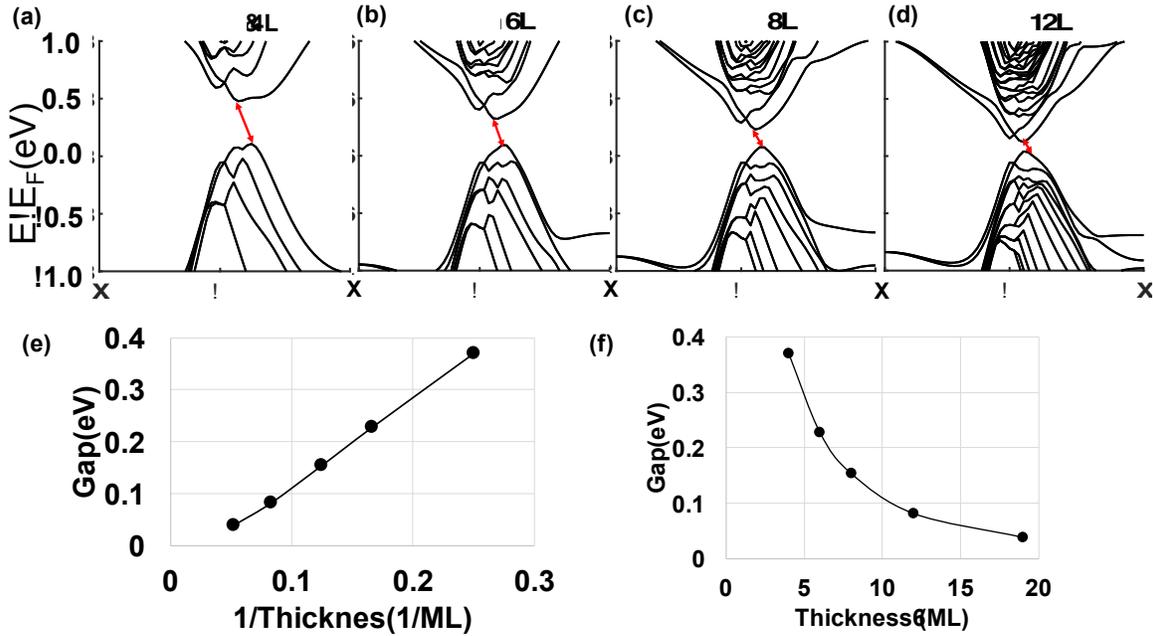

**Figure S.I. 9:** (a) – (d) Band structures of defect-free SnTe slabs for 4, 6, 8, 12 ML thicknesses show how the hybridization gap decreases with increasing thickness. Plots illustrating this fact show the hybridization gap as a function of (e) inverse of thickness and (f) thickness of the slab.

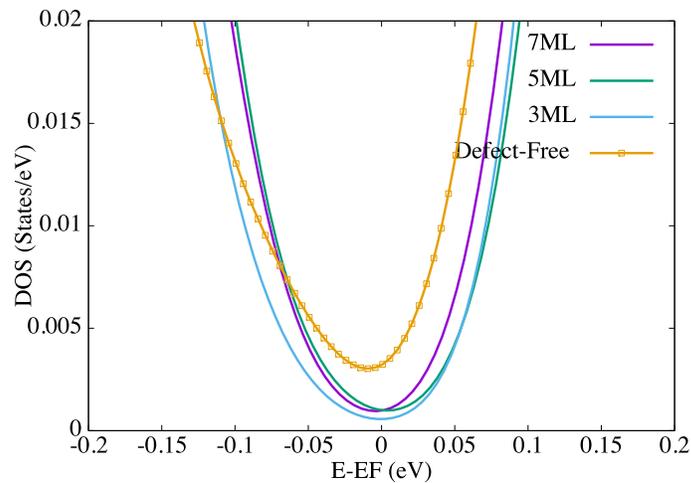

**Figure S.I. 10:** Surface layer density of states (DOS) of slabs with thickness of 8 ML and terrace sizes of 3, 5 and 7 ML (vicinal surfaces). The surface DOS for a defect free slab with the similar thickness is presented as well.

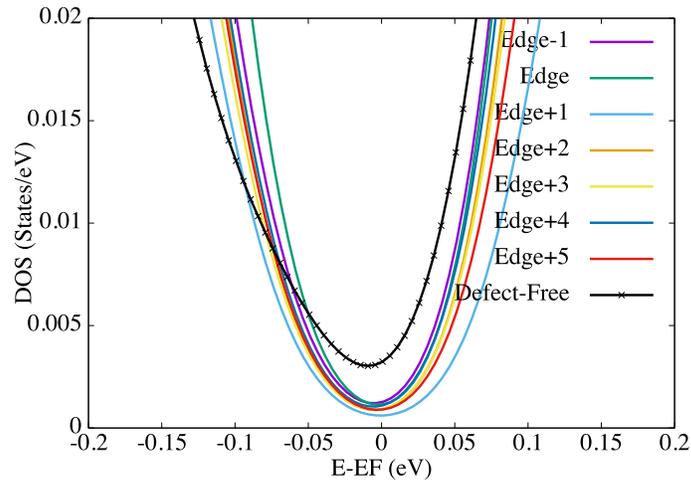

**Figure S.I. 11:** Projected surface density of states (DOS) on different atomic rows (parallel to the step edges) on the terrace of a vicinal stepped surface with slab thickness of 8 ML and terrace sizes of 7 ML. The corresponding DOS for a defect free slab with the similar thickness is presented.

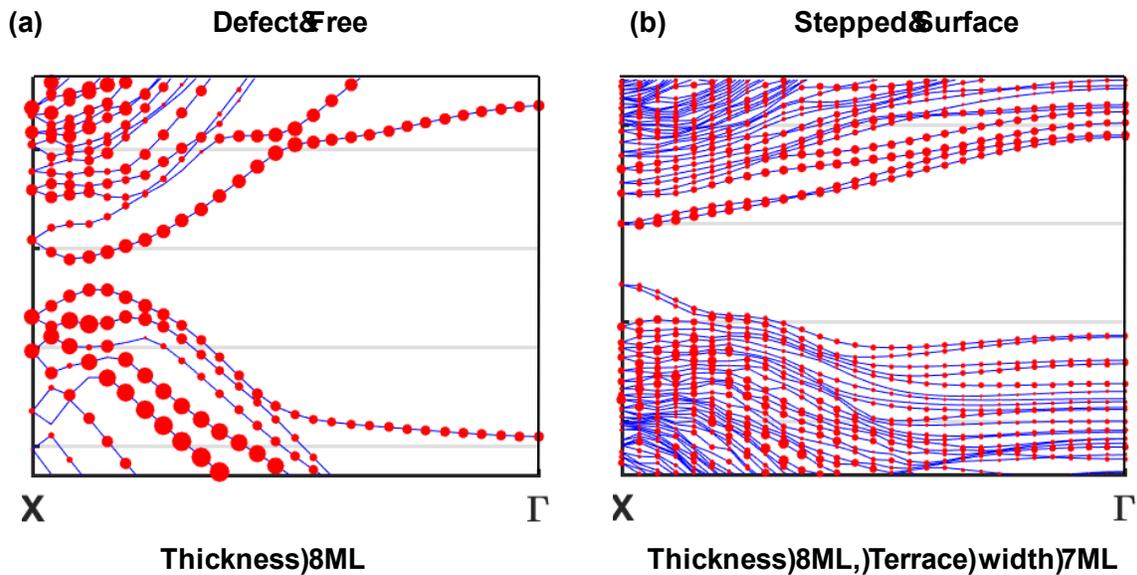

**Figure S.I. 12:** Band structure for defect-free (a) and stepped surfaces (b); blue curves are computed for finite slabs with 8 ML thickness. The size of the red circles indicates the total projection of each band on to the surface atoms. The SS exist in both cases but have greatly reduced weight on the surface of the stepped surface. The small gap (i.e., lack of Dirac

linear band crossing) for the SS is due to the finite slab thickness as described in Fig. S.I. 9.

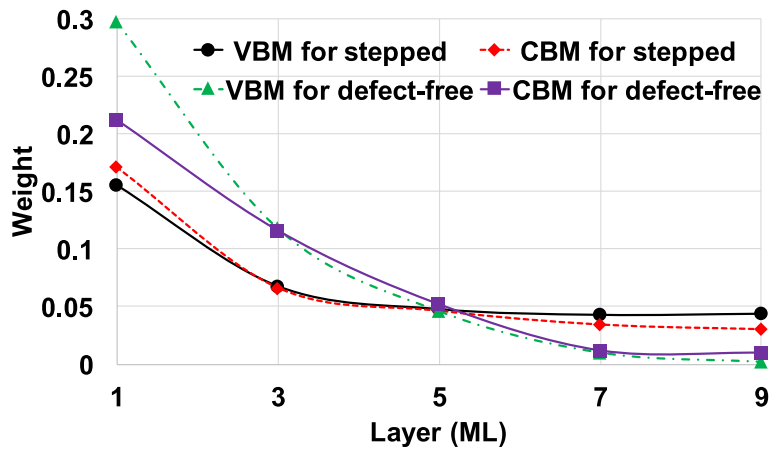

**Figure S.I. 13:** Computed total SS weight on each monolayer as a function of layers going into the interior of the slab for both the defect-free (green and purple) and stepped surfaces (black and red) for a k-point near Dirac point. The bands chosen are the valance band maxima (VBM) and conduction band minima (CBM).

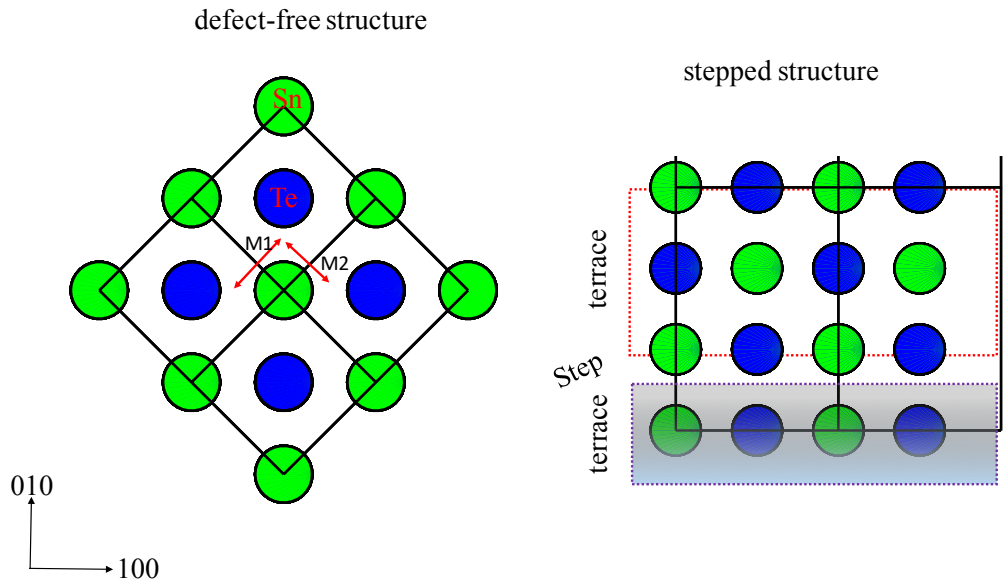

**Figure S.I. 14:** Two rotational symmetries are shown as double-headed arrow on the left (M1 and M2) for the top view of the SnTe (001) surface on the left. On the right, this top view shows that we break both symmetries across the surface step when going from one

terrace to the next; the lower shaded area (blue dashed box) indicates a separate terrace from the highlighted terrace (red dashed box).

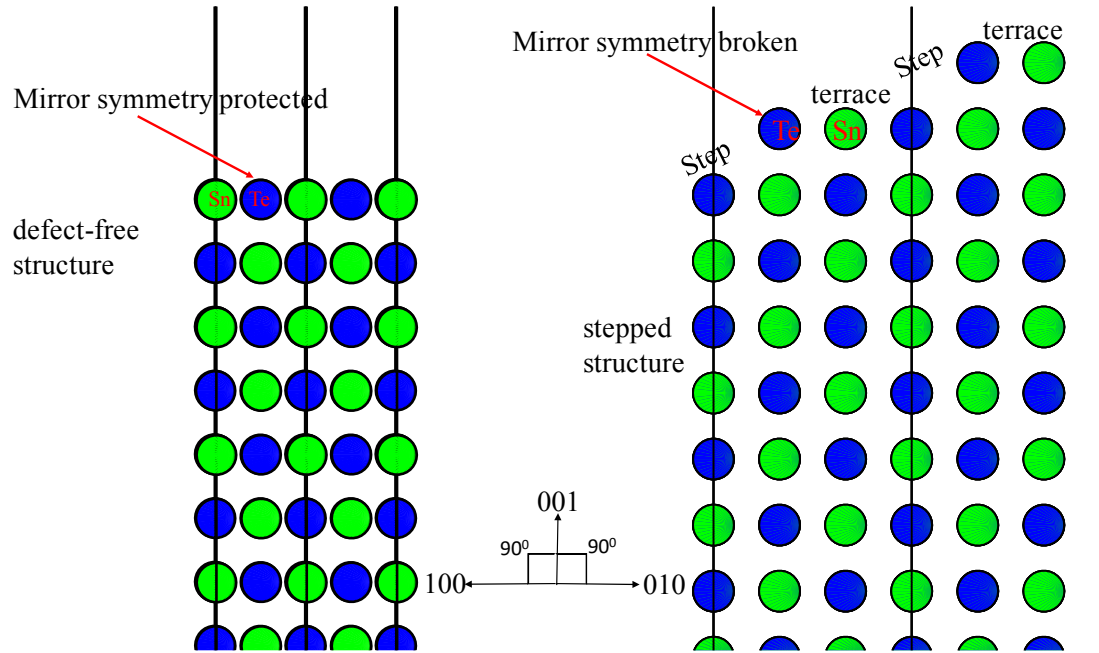

**Figure S.I. 15:** Side-views of the unit cell for both defect-free (left) and stepped (right) structures. Red arrows indicate where the rotational symmetry is protected (left) and broken (right).

**References for Supplemental Material**

7. C.M. Polley, V. Jovic, T.Y. Su, M. Saghir, D. Newby, B.J. Kowalski, R. Jakiela, A. Barcz, M. Guziewicz, T. Balasubramanian, G. Balakrishnan, J. Laverock, and K.E. Smith, Observation of surface states on heavily indium-doped SnTe(111), a superconducting topological crystalline insulator. Physical Review B **93**, 075132 (2016).
8. O.E. Dagdeviren, C. Zhou, K. Zou, G.H. Simon, S.D. Albright, S. Mandal, M.D. Morales-Acosta, X. Zhu, S. Ismail-Beigi, F.J. Walker, C.H. Ahn, U.D. Schwarz, and E.I. Altman, Length Scale and Dimensionality of Defects in Epitaxial SnTe Topological Crystalline Insulator Films. Advanced Materials Interfaces **4**, 1601011-10 (2017).
9. N.W.a.M. Ashcroft, N.D., *Solid State Physics*. 1981, Philadelphia: Sounders College.
10. L.E. Davis, N.C. MacDonald, P.W. Palmberg, G.E. Riach, and R.E. Weber, *A Refence Book of Standard Data for Identification and Interpretation of Auger Electron Spectroscopy Data*. 2nd ed. 1976, Minnesota, USA: Physics Electronics Industries Inc.
11. D. Zhang, H. Baek, J. Ha, T. Zhang, J. Wyrick, A.V. Davydov, Y. Kuk, and J.A. Stroscio, Quasiparticle scattering from topological crystalline insulator SnTe (001) surface states. Physical Review B **89**, 245445 (2014).
12. X. Qian, L. Fu, and J. Li, Topological crystalline insulator nanomembrane with strain-tunable band gap. Nano Research **8**, 967-979 (2014).
13. S. Liu, Y. Kim, L.Z. Tan, and A.M. Rappe, Strain-Induced Ferroelectric Topological Insulator. Nano Letters **16**, 1663-1668 (2016).